\documentclass{appolb}

\usepackage{graphicx}
\usepackage{wasysym}
\usepackage{amssymb}

\newcommand{\rin}{r_\mathrm{in}}
\newcommand{\rout}{r_\mathrm{out}}
\newcommand{\mc}{M_\mathrm{c}}
\newcommand{\md}{M_\mathrm{d}}
\newcommand{\ms}{M_{\astrosun}}

\begin{document}

\title{Estimating masses of Keplerian disk systems:\\ the case of AGN in NGC 4258}
\author{Patryk Mach, Edward Malec and Micha\l~Pir\'og
\address{M. Smoluchowski Institute of Physics, Jagiellonian University, Reymonta 4, 30-059 Krak\'ow, Poland}}

\maketitle

\begin{abstract}
\tolerance=500
The Keplerian motion of accretion disks in Active Galactic Nuclei (AGN) is usually believed to be generated by a heavy central mass. We investigate accreting disk systems with polytropic gas in Keplerian rotation and obtain a phenomenological formula that relates the Keplerian angular frequency to the ratio of disk and central masses. Central mass approaches the Keplerian value, if the inner boundary of a disk is close to the minimal stable orbit of a black hole. These results are applied to NGC 4258, the unique AGN with a finely measured Keplerian rotation curve of the central disk, with the conclusion that its rotation curve is, in fact, determined by the central black hole. The mass of the accretion disk exceeds $100\ms$.
\end{abstract}

\vspace{1em}

In the astrophysical literature, the velocity of rotation of disks in Keplerian motion is typically associated with a heavy central mass $\mc$ and a light disk mass $\md$. The discovery of the massive black hole in the masing nucleus of NGC 4258 exemplifies this interpretation \cite{Miyoshi1995}. On the other hand, it is known that even the selfgravity of heavy disks can be compatible with the Keplerian rotation \cite{Hashimoto1995}. Hur\'{e} et al.~\cite{Hure2011} investigated thin dust disks and pointed out that the rotational Keplerian velocity gives only information on the enclosed mass of the system, and it is not possible to separate masses of the two components --- of a disk and a central mass.

We show in this paper that the Keplerian rotation curve, with the frequency $\omega $, does depend on the two masses but they do not contribute additively. A phenomenological formula expresses $\mc G/(r^3\omega^2)$ ($r$ and $G$ are the cylindrical radius and the gravitational constant, respectively) as a function of $x\equiv \md /\mc $ and of the ratio $y\equiv \rin /\rout $, where $\rin $ and $\rout $ are the innermost and outermost disk radii, respectively. If $y$ is small enough then the central mass is reasonably well approximated by the Keplerian value $r^3\omega^2/G$.

Consider a stationary, axially symmetric, selfgravitating disk of gas, rotating around a central point mass $\mc$. The gravitational potential of the system can be written as the superposition $\Phi = - G \mc/|\mathbf x| + \Phi_\mathrm{g}$. The potential $\Phi_\mathrm{g}$ is due to the gravity of the disk and satisfies the Poisson equation
\begin{equation}
\label{poisson}
\Delta \Phi_\mathrm{g} = 4 \pi G \rho,
\end{equation}
where $\rho$ denotes the mass density of the gas. In the following, we use the cylindrical coordinates $(r,\phi,z)$. The velocity of the gas reads $\mathbf U = \omega(r,z) \partial_\phi$. For a barotropic equation of state the frequency $\omega$ depends only on the cylindrical radius $r$ \cite{Tassoul1978}. We assume polytropic equations of state $p = K\rho^\Gamma $, where $p$ is the gas pressure, and $K$ and $\Gamma$ are constant. The Euler equations 
\[ \nabla p + \rho (\mathbf U \cdot \nabla) \mathbf U + \rho \nabla \Phi = 0 \]
can be integrated, yielding
\begin{equation}
\label{euler}
h + \Phi_\mathrm{c} + \Phi = C
\end{equation}
in the region where $\rho \neq 0$. Here, $h$ denotes the specific enthalpy of the fluid $dh = dp/\rho$, and
\[ \Phi_\mathrm{c} = - \int^r dr^\prime r^\prime \omega^2 \left( r^\prime \right) \]
is the centrifugal potential. The structure of the disk can be obtained from Eqs.~(\ref{poisson}) and (\ref{euler}) provided that the equation of state and the rotation law $\omega = \omega(r)$ are known.
 
Let the Keplerian rotation law be $\omega = \omega_0/r^{3/2}$. It follows from Eq.~(\ref{euler}) that
\begin{equation}
\label{omega}
\omega_0^2 = GM_\mathrm{c} + \frac{r_\mathrm{in} r_\mathrm{out} \left( \Phi_\mathrm{g}(r_\mathrm{out}) - \Phi_\mathrm{g}(r_\mathrm{in}) \right)}{r_\mathrm{out} - r_\mathrm{in}},
\end{equation}
where $\Phi_\mathrm{g}(r_\mathrm{in}) = \Phi_\mathrm{g}(r=r_\mathrm{in},z=0)$ and $\Phi_\mathrm{g}(r_\mathrm{out}) = \Phi_\mathrm{g}(r=r_\mathrm{out},z=0)$. We fix $\rout$ and the maximal density $\rho_\mathrm{max}$. Then, we find that the limit of the right-hand side of Eq.~(\ref{omega}) for $r_\mathrm{in} \to 0$ is the Keplerian angular frequency $\omega_0^2 = G M_\mathrm{c}$. To show that, note that $\Phi_g$ is a bounded function on $\mathbb R^3$. It is also possible to prove that $\omega_0^2 \to G \mc$ as $\rin \to \rout$.

\begin{figure}[t]
\begin{center}
\includegraphics[width=0.9\textwidth]{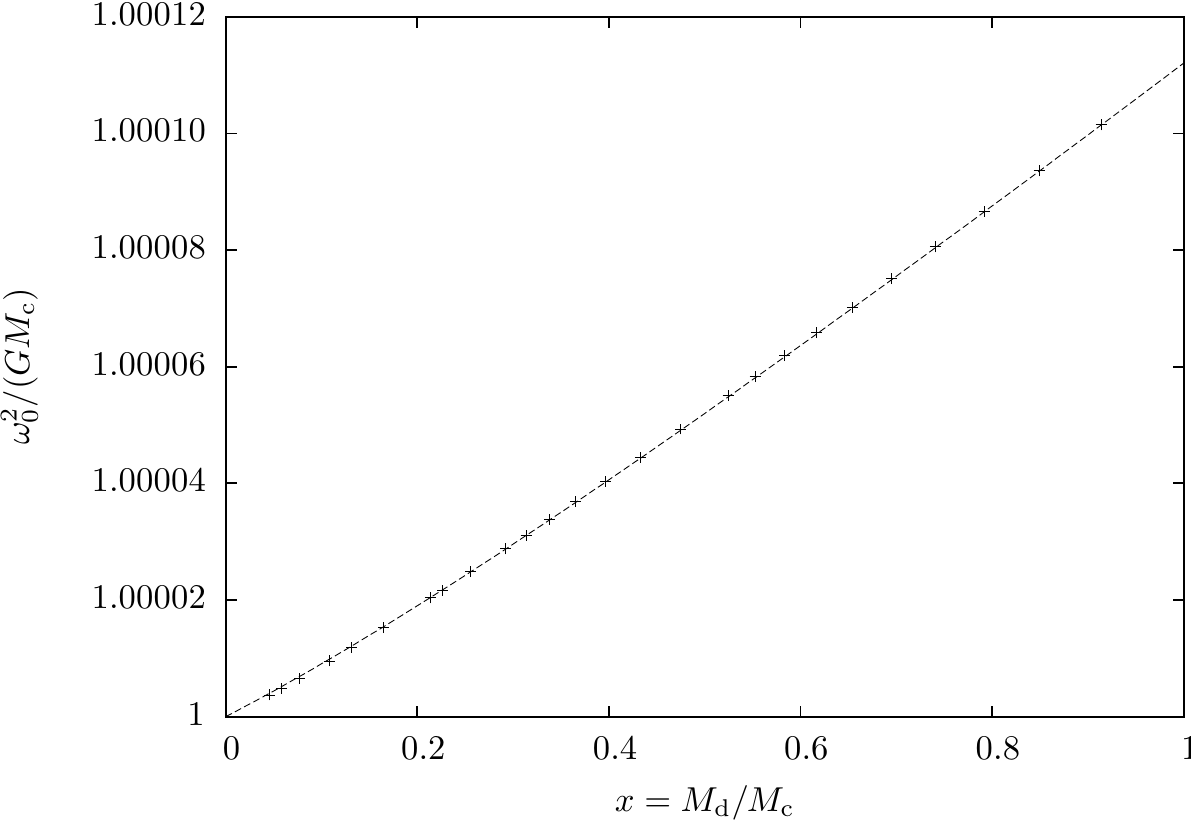}
\end{center}
\caption{\label{fig:1} The dependence of $\omega_0^2/(G \mc)$ on the mass ratio $x$. The plot is obtained for the adiabatic index $\Gamma = 5/3$ and $\rin/\rout = 10^{-4}$. The dashed line depicts analytic fit (\ref{omega-fit}). Crosses correspond to numerical solutions obtained for different values of $\mc$. }
\end{figure}

\begin{figure}[t]
\begin{center}
\includegraphics[width=0.9\textwidth]{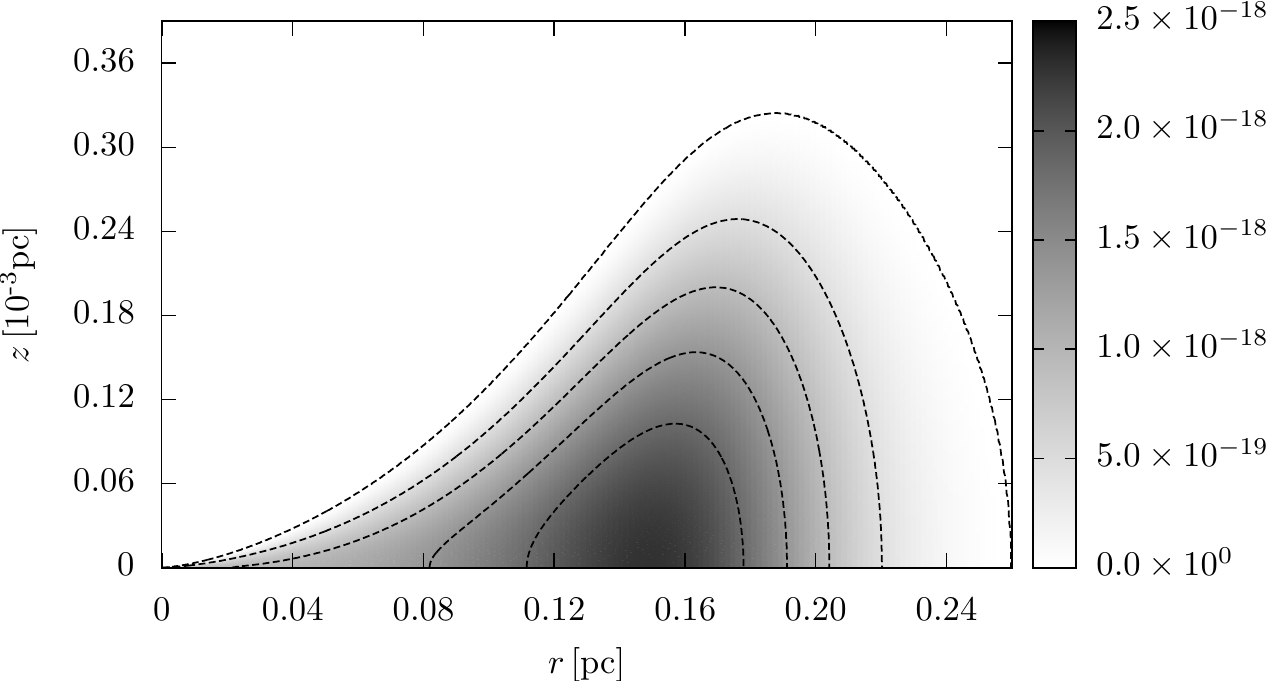}
\end{center}
\caption{\label{fig:2}Mass density profile within a disk model obtained for $\Gamma = 5/3$. The units are $\mathrm{g \, cm^{-3}}$.}
\end{figure}

In the numerical analysis we apply the Self-Consistent Field (SCF) scheme \cite{Ostriker1968, Clement1974, Blinnikov1975}. Numerical solutions converge quite well for Keplerian rotation, although the drawback of the method is the need to employ a large number $l$ of Legendre polynomials for thin disks. We find accurate solutions with $ l=400$ and large grids (up to $5000 \times 5000$). Such resolutions are permitted because of the simplicity of the SCF method. Our numerical solutions have been tested by the new version of the virial theorem that includes a point mass \cite{mach_virial}. It should be pointed out, however, that even $l=400$ does not suffice to resolve structures in the gravitational potential that are thinner than thinnest disks considered in this paper. We performed tests on thicker disks, demonstrating that important quantities are insensitive to the addition of a higher number of expansion polynomials, which could model fine structures of the gravitational potential within the disk. This conclusion is also supported by the heuristic argument provided by the analysis of infinitely thin disks --- they give a regular gravitational potential, i.e., without any structure comparable with the disk thickness.

We found a few thousands of numerical solutions for polytropic disks, mainly with two polytropic indices $\Gamma = 5/3$ (which can be thought to represent monoatomic gas) and $\Gamma = 4/3$ (which might well be associated with vapor water maser disks) and for a few dozens of values of the interior radius corresponding to $y\in \left(10^{-4}, 0.5\right)$. Calculations have been done also for a range of polytropic indices in between 4/3 and 5/3, and the internal radius corresponding to $y =10^{-4}$. 

Below we shall report results concerning the case $\Gamma = 5/3$, but the essential features of solutions are similar in other cases. The obtained results can be quite well approximated in the quadrant $0 < x < 1$, $10^{-5}< y < 0.55$ by the following phenomenological formula
\begin{equation}
\label{omega-fit}
\frac{\omega_0^2}{G \mc} = 1 + x f \left( y\right) \frac{ g \left( y\right) + x }{h\left(y\right) + x},
\end{equation}
where $x \equiv \md / \mc$ and $f$, $g$ and $h$ are defined as follows
\begin{eqnarray*}
f(y) & = & 1.322y - 3.719y^2 + 13.47y^3 - 44.48y^4\\
     &   & +116.9y^5 -206.2y^6 + 208.9y^7 - 90.60y^8,\\
g(y) & = & 0.4778 + 0.9495y - 31.67y^2 + 258.4y^3\\
     &   & -1431y^4 + 5169y^5 - 11100y^6 + 12760y^7 -6034y^8,\\
h(y) & = & 0.7423 + 0.6313y - 43.57y^2 + 375y^3\\
     &   & -2059y^4 + 7281y^5 - 15390y^6 + 17520y^7 -8242y^8.
\end{eqnarray*}
Functions $f$, $g$ and $h$ are nonnegative and smaller than 1 for arguments $0\le y \le 1$. 

It is clear from these formulae that the central mass is always smaller than $\omega^2 r^3/G$. At the same time, it is easy to see from (\ref{omega-fit}) that if $y$ is sufficiently small and $\md < \mc $, then the rotation curve of the gaseous disk is almost exclusively influenced by the central mass.

The number of known AGN's with masers is increasing in recent years \cite{Braatz1997, Greenhill1997, Gwinn1997, Trotter1998, Yamauchi2004, Kondratko2005, Ishihara2001, Greenhill2003} (see also for a review \cite{Kondratko2008}), but there there is only one case of well resolved Keplerian rotation curve --- in the AGN of NGC 4258. Observations of supermasers near the center of the galaxy NGC 4258 \cite{Miyoshi1995, Moran1995} provide evidence for a rotating disk of gas surrounding a black hole with a mass $\mc = (3.9 \pm 0.3) \times 10^7 \ms$. The motion is Keplerian to high precision, of about a part in a hundred \cite{Miyoshi1995, Maoz, Moran2008, Herrnstein2005}, and this mass estimate assumes that velocities are due solely to the gravitational force of the central body. A similar result has been obtained by Siopis et al.~\cite{Siopis2009}, who determined the mass of the black hole by constructing axisymmetric dynamical models of the galaxy NGC 4258. Their best mass estimate yields $M_\mathrm{c} = (3.3 \pm 0.2) \times 10^7 \ms $.

The masers are observed in the annular region that extends from 0.13 pc to 0.26 pc from the center, and the region with detected masers is thin --- the maximal relative height is circa 1/800 (these distances can be expressed in terms of the Schwarzschild radius: $r_\mathrm{g} \equiv G \mc / c^2 = 5.8 \times 10^{12} $cm and $1 \mathrm{pc} = 5.4 \times 10^5 r_\mathrm{g}$). There exists evidence that the disk fills not only the annular region in which masers are detected, but that it extends outwards from a vicinity of the black hole horizon. Reynolds et al.~\cite{Reynolds2009} investigated the weakly radiating AGN in NGC 4258 using data from Suzaku, XMM-Newton, and the Swift/Burst Alert Telescope survey. They constrained its luminosity region to between 10 and $4 \times 10^4$ Schwarzschild radii of the black hole. In the numerical calculations we assume that the disk extends from $\rin = 2.6 \times 10^{-5}$ pc to $\rout =0.26$ pc, and its relative height is equal to 1/800.

It is clear from (\ref{omega-fit}) --- see also Fig.~1 --- that if $\md \le \mc $ then $\omega_0^2/(G \mc) \approx 1$ with the deviation from unity much smaller than 0.001. The disk's mass, obtained numerically for the polytropic index $\Gamma = 5/3$ and assumed geometrical parameters of the disk, reads $\md = 2.62 \times 10^{-6} \mc \approx 102 \ms $.

The results for other polytropes, $5/3> \Gamma \ge 4/3$, suggest that disk's mass depends rather weakly on the equation of state and increases (up to $\md =190 \ms $) with the decrease of the polytropic index. The disk's size can be larger than that detected in observations, which, in turn, would yield a larger mass. Therefore $\md =100 \ms $ can be regarded as a conservative lower bound on the mass of the gaseous disk in the AGN of NGC 4258. A sample distribution of the mass density within the disk, obtained for $\Gamma = 5/3$, is shown on Fig.~2.

The above results are consistent with those rough mass estimates that base on such values of the baryonic number density $\rho_\mathrm{n}$ within the disk that favor the existence of water masers. If one assumes that $\rho_\mathrm{n}$ is around $10^{10}$ per $\mathrm{cm}^3$, then the mass is around $10^4 \ms$ \cite{Miyoshi1995, Moran1995}. Herrnstein et al.~\cite{Herrnstein2005} note that the allowed density is $10^8/\mathrm{cm}^3 < \rho_\mathrm{n} < 10^{10}/\mathrm{cm}^3$; the lower bound gives the disk mass $\approx 100 \ms$.

This analysis is entirely Newtonian, and since it applies to systems with black holes, it should be repeated in the general-relativistic context. We do not expect essential changes concerning those systems, where the inner disk boundary is well separated from a black hole horizon, $\rin \gg 6 \mc$.

\section*{Acknowledgments}

The research was carried out with the supercomputer ``Deszno'' purchased thanks to the financial support of the European Regional Development Fund in the framework of the Polish Innovation Economy Operational Program (contract no. POIG.02.01.00-12-023/08).

\end{document}